# Design Method for Constant Power Consumption of Differential Logic Circuits

Kris Tiri[1] and Ingrid Verbauwhede[1,2]

[1]UC Los Angeles, [2]K.U.Leuven
{tiri,ingrid}@ee.ucla.edu

**Abstract**

*Side channel attacks are a major security concern for smart cards and other embedded devices. They analyze the variations on the power consumption to find the secret key of the encryption algorithm implemented within the security IC. To address this issue, logic gates that have a constant power dissipation independent of the input signals, are used in security ICs. This paper presents a design methodology to create fully connected differential pull down networks. Fully connected differential pull down networks are transistor networks that for any complementary input combination connect all the internal nodes of the network to one of the external nodes of the network. They are memoryless and for that reason have a constant load capacitance and power consumption. This type of networks is used in specialized logic gates to guarantee a constant contribution of the internal nodes into the total power consumption of the logic gate.*

## 1 Introduction

Encryption algorithms are designed to be secure against cryptanalysis that has access to plaintext and ciphertext. The physical implementation however, provides the attacker with important information. Numerous attacks have been presented that use 'side-channels', such as the time delay and the power consumption of the encryption operation, as an extra source of information to find the secret key [1]. These so-called side-channel attacks threaten any device of which the integrated circuit is easily observable such as smart cards and embedded devices. The differential power analysis [2] in particular is of great concern. It is very effective in finding the secret key and can be mounted quickly with off-the-shelf devices. The attack is based on the fact that logic operations have power characteristics that depend on the input data. It relies on statistical analysis to extract the information from the power consumption that is correlated to the secret key.

All embedded applications, such as for example PDAs, cellphones, smart cards and sensor nodes in an ambient intelligent environment need to communicate with their environment. Therefore, they will need cryptographic capabilities for authentication and confidentiality. Some efforts have recently been made during special sessions at premier design conferences to raise the awareness for the danger that side-channel attacks pose to all embedded devices [3],[4].

At first, the differential power analysis attack has been fought with ad hoc countermeasures. For instance, the addition of random power consuming operations or a current sink obscured the data dependent variations in the power consumption. These techniques do not solve the problem; they only increase the number of measurements for a DPA attack. Subsequently, countermeasures have been conceived at different abstraction levels of the security application. One illustration at the algorithmic level is masking [5]. This technique prevents that intermediate variables depend on an easily accessible subset of the secret key.

However, the best solution is to try not to *create* any side channel information. This can be done at the logic level with specialized circuit techniques [6],[7] The goal of these countermeasures is to make the power consumption of the individual logic gates independent of their input signals or the Hamming distances between subsequent input signals. When the power consumption of the smallest building block is a constant and independent of the signal activity, no information is leaked through the power supply and power attacks are impossible.

The circuit techniques can be categorized in two groups. One group uses gates composed of standard logic gates [8],[9], the other group uses custom made gates [10],[11],[12]. Custom made gates offer a higher security because the designer has full control over the internal organization of the cells. Yet, there is a tradeoff with the costs of developing a custom designed library. In dynamic current mode logic (DyCML) [13], proposed in [11], the transistor sizes even depend on the final layout of the interconnect wiring after place & route as they are based on the output capacitance. DyCML also requires a complex clock delay network as gates at different logic depth need different clock delays. Major disadvantages of the logic style proposed in [12] are the limited library – only the AND-NAND gate–, the large transistor count – 112 transistors for the AND-NAND gate– and the asynchronous design approach, which still lags behind that of synchronous designs.

For sense amplifier based logic (SABL), our previous work proposed in [10], only gates with two or fewer inputs have been presented. The logic style requires a special structure of the differential pull down network (DPDN) to control the contribution of the internal nodes into the power consumption. The main requirement is that the network must be fully connected. This means that for any complementary input combination all the internal nodes of the network connect to one of the external nodes of the network. As a result, all the internal nodes go through a discharge-charge cycle and have a constant power dissipation. The main contribution of this paper is a systematic design method to generate fully connected differential pull down networks for an arbitrary logic function.

The remainder of this paper is organized as follows. The next section enumerates the characteristics of constant power dissipating logic. Section 3 discusses the fully connected differential pull down network in detail. In section 4, design methods are formulated to create fully connected differential pull down networks given a Boolean expression or given an existing differential pull down network. The design example implements a complex fully connected differential pull down network. Section 5 describes an enhancement to avoid early propagation. Finally, a conclusion will be formulated.

## 2 Constant Power Dissipating Logic

The power dissipation of traditional logic depends on the signal activity. When the output of the logic gate makes a 0 to 1 transition, a current comes from the supply and charges the output capacitance. On the other hand, when the output sees a 1 to 0 transition or no transition at all, no energy is consumed from the power supply. Two conditions must be satisfied to have constant power dissipating logic: (1) a logic gate must have exactly one charging event per clock cycle; and (2) the logic gate must charge a constant capacitance in that charging event.

Dynamic and differential logic fulfills the first condition. In a differential logic family, a signal is represented by both the true and the false value. A dynamic logic family alternates precharge and evaluation phases. In the precharge phase, the output is pre-charged to 1, and in the evaluation phase the output is conditionally evaluated to 0. For the combination of dynamic and differential logic this means that exactly one of the two outputs will evaluate to 0 as the output must be differential during the evaluation phase. During the subsequent precharge phase, the discharged output will be pre-charged to 1. In other words, every signal transition, including the events in which the input signals remain constant, is represented with an actual switching event, in which the logic gate charges a capacitance. All the logic families that have been introduced to thwart the DPA [7],[8],[9],[10],[11],[12] employ some form of dynamic differential logic.

Besides a 100% switching factor, it is essential in order to achieve constant power consumption that a fixed amount of charge is used per transition. Differential logic has a load capacitance at each output. Since only one output undergoes a transition per switching event, the total load at the true output should match the total load at the false output. This means that the load capacitances at the differential output should be matched. The load capacitance has three components: the intrinsic output capacitance of the gate, the interconnect capacitance and the intrinsic input capacitance of the load. Intrinsic capacitances are parasitic node capacitances of the logic gate.

(Dis)charging parasitic capacitances, which are internal to the gate itself, have an influence on the power consumption as well. Simulations indicate that e.g. for the AND-NAND gate in cascode voltage switch logic (CVSL) [14], the variation on the power consumption can be as large as 50% [10]. This is caused by asymmetry in the gate. Depending on the input, different parasitic capacitances that are internal to the differential pull down network discharge during the evaluation phase. In the succeeding power consuming precharge phase, these capacitances are recharged. This phenomenon is also referred to as memory effect.

Sense amplifier based logic [10] uses advanced circuit techniques to guarantee that the load capacitance has a constant value. The logic style completely controls the portion of the load capacitance that is due to the logic gate. The intrinsic capacitances at the differential in- and output signals are symmetric. Additionally, the special structure of the differential pull down network guarantees that all the internal nodes go through a discharge-charge cycle and no nodes are left floating. As a result it discharges and charges each of the internal nodes in every cycle together with one of the balanced output capacitances. Hence, it discharges and charges a constant capacitance value.

The generic sense amplifier based logic gate is shown in Fig. 1. The gate consists of the sense amplifier of the StrongArm110 flip-flop [15] of which the input differential pair is exchanged by a differential pull down network. In the precharge phase, both the true and the false value of all input signals are set at 0. The true and the false output signal are precharged to 1. During the evaluation phase, the input becomes complementary and the network connects one of the network output nodes, X or Y, to the common node Z. This will make the logic gate switch and one gate output will discharge to 0. Note that whichever branch of the differential pull down network is on, X and Y are connected through transistor $M_1$ and both nodes will eventually be discharged. This leaves the gate in a stable state until the next precharge phase.

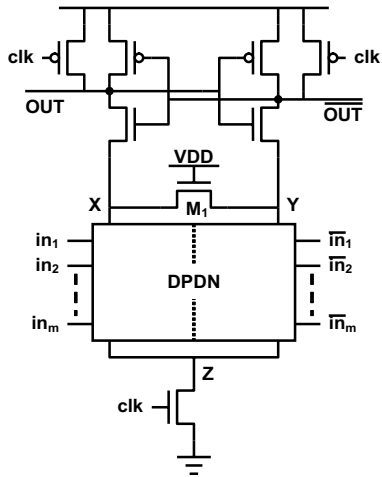

**Fig. 1. Generic sense amplifier based logic gate.**

## 3 Fully Connected DPDN

A fully connected differential pull down network is designed such that for any complementary input combination all internal nodes of the network are connected to one of the module output nodes X or Y. As a result, since both nodes X and Y eventually discharge, all the internal nodes are discharged during the evaluation phase and will be charged during the next precharge phase. The requirement of fully connected differential pull down networks in constant power dissipating logic is in contrast with genuine logic, where the design constraints are to minimize the device count and the number of stacked levels [16].

As an example, Fig. 2 (left) and (right) show the implementation of the AND-NAND function with a genuine and with a fully connected differential pull down network respectively. The genuine network has one internal node, node W. This node can become floating for some input combinations. For example when both A and B remain at 0 during the transition from the precharge phase to the evaluation phase (and $\overline{A}$ and $\overline{B}$ both switch from 0 to 1), node W is disconnected from node X as well as from node Z. There is no path for the charge to flow away and node W keeps it charge on the parasitic capacitance. On the other hand, when both A and B switch from 0 to 1 ($\overline{A}$ and $\overline{B}$ both remain at 0), node W is connected to node Z and the parasitic capacitance discharges. The difference in power dissipation between both input events will come from the fact that for the latter event, the capacitance at node W will need to be charged in the subsequent precharge phase.

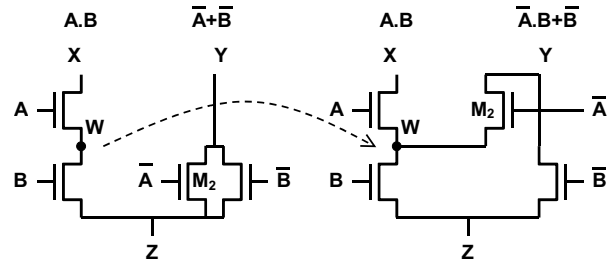

**Fig. 2. Implementation of AND-NAND function with genuine DPDN (left); and fully connected DPDN (right).**

For a genuine differential pull down network several discharge events with different combinations of parasitic capacitances exist. This means that the total load capacitance, and thus the power consumption, is signal dependent. It is impossible to match the total load capacitance of the different discharge events by manipulating transistor sizes and/or layouts. The contribution of the internal node capacitances can only be controlled by making sure that they all discharge independently of the input event. Repositioning transistor $M_2$, which is located between nodes Y and Z and is driven by input $\overline{A}$, between nodes Y and W achieves this effect. This operation does not alter the functionality of the individual branches of the differential pull down network ($A.B = A.B$ and $\overline{A}+\overline{B} = \overline{A}.B+\overline{B}$) but it guarantees that for any complementary input combination, node W is connected to one of the module output nodes. Hence, independently of the input event in the evaluation phase, node W will be discharged.

Fig. 3 shows two transient SPICE simulations of a discharging event in the evaluation phase followed by a charging event in the precharge phase of the sense amplifier based logic AND-NAND gate for two different inputs. The figure shows that the instantaneous output voltages and the supply current are independent of the input event applied to the gate. Fig. 4 demonstrates that for both events all the internal node capacitances and one of the balanced output nodes are discharged. In each event, the same amount of charge is needed to charge the same capacitances, and hence the same amount of energy, is used.

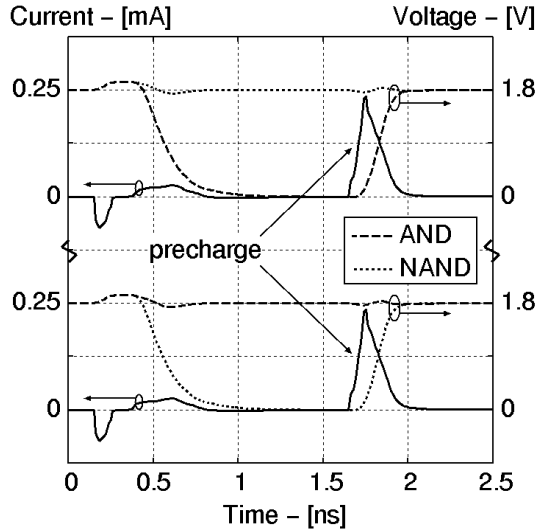

**Fig. 3. Output signals and supply current of SABL AND-NAND gate: transient simulation for (0,1)-input (top); and (1,1)-input (bottom). Note that in the precharge phase all inputs (A, $\bar{A}$, B, $\bar{B}$) are at 0 and that subsequently in the evaluation phase either A or $\bar{A}$ and either B or $\bar{B}$ become 1.**

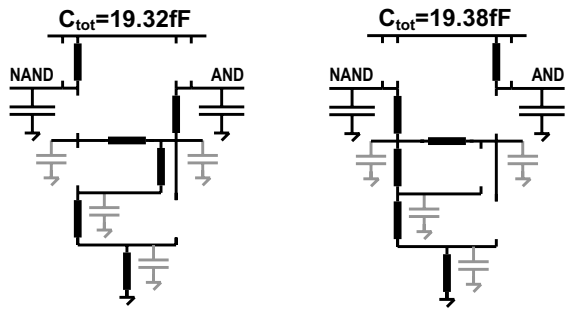

**Fig. 4. Discharge event of SABL AND-NAND gate in the evaluation phase for (0,1)-input (left); and (1,1)-input (right).**

## 4 Designing fully connected DPDNs

This section formulates a design method to create a fully connected differential pull down network for a logical function $f$ given either a Boolean expression or an existing differential pull down network.

### 4.1 Given a Boolean expression

The design procedure to create a fully connected differential pull down network for a logical function $f$ consists of five steps.

Step 0: Create the Boolean expression of the logical function $f$. The complementary output will be referred to as $\bar{f}$.

Step 1: Identify 2 expressions $x$ and $y$ that combine to the logical function $f$. The result is either an AND-operation ($f = x \cdot y$) or an OR-operation, ($f = x + y$).

Step 2: Complement the expression of $f$ in $x$ and $y$ to get the dual expression $\bar{f}$. The result is an OR-operation ($\bar{f} = \bar{x} + \bar{y}$) or an AND-operation, ($\bar{f} = \bar{x} \cdot \bar{y}$).

The results of the previous 2 steps are two dual expressions:

either case A)      or case B)
$$\begin{cases} f = x \cdot y \\ \bar{f} = \bar{x} + \bar{y} \end{cases} \qquad \begin{cases} f = x + y \\ \bar{f} = \bar{x} \cdot \bar{y} \end{cases}$$

Step 3: Transform the OR-operation.

The Boolean expression is translated to a differential PDN in the traditional way. An AND operation is represented by a series of switches, an OR operation by a parallel connection of switches [17]. At this abstraction level, only the series combination has an internal node.

In case A), we transform the parallel connection into $\bar{x} \cdot y + \bar{y}$, put network $y$ at the bottom of the $x \cdot y$ connection and share network $y$ between the two branches $x \cdot y$ and $\bar{x} \cdot y + \bar{y}$.

In case B) we transform the parallel connection into $x \cdot \bar{y} + y$, put network $\bar{y}$ at the bottom of the $\bar{x} \cdot \bar{y}$ connection and share network $\bar{y}$ between the two branches $\bar{x} \cdot \bar{y}$ and $x \cdot \bar{y} + y$.

Step 4: Decompose and repeat the procedure for the logical expressions $x$ and $y$ until the network consists of only 1 literal, which corresponds to a single transistor.

Step 5: Substitute the results.

### 4.2 Given an existing DPDN

The above design method can also be implemented directly on a differential pull down network. For a given schematic of a genuine differential pull down network, the design procedure is a transformation. The transformation repositions transistors in the genuine network in three steps. The total number of devices remains the same between the genuine and the fully connected network. The evaluation depth, which is defined by the maximum number of devices in series, may increase.

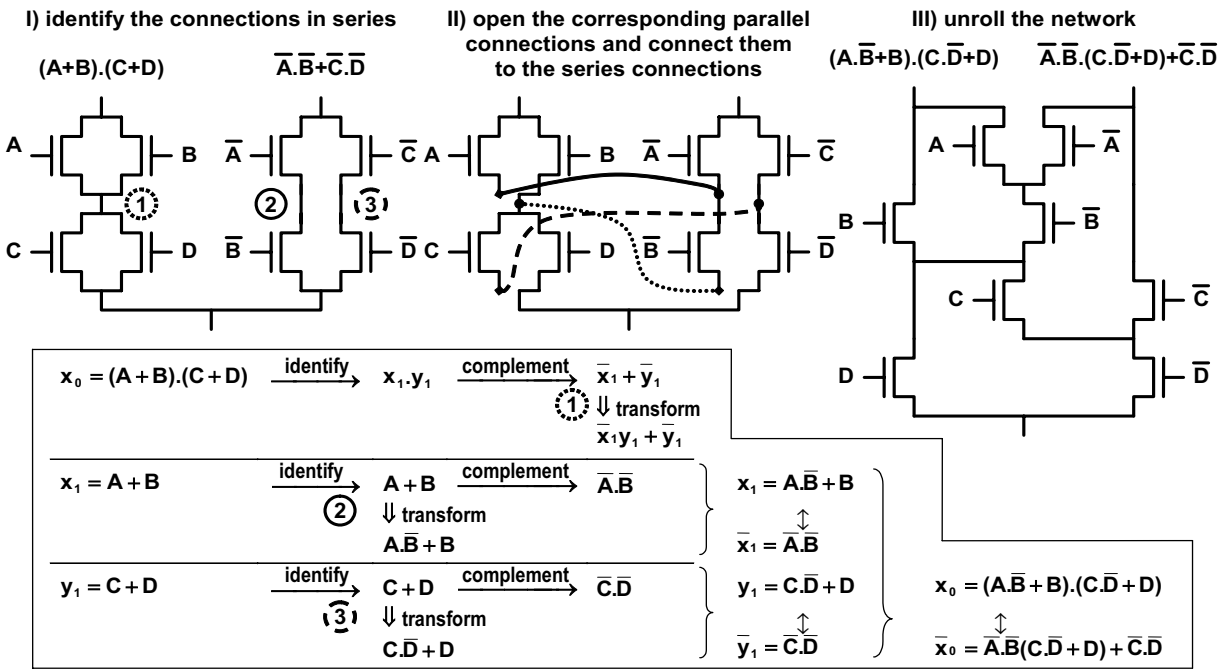

**Fig. 5.** Transformation of a complex DPDN to a fully connected DPDN: design example.

Step 1: Identify all the networks in series.

Step 2a: Open the corresponding dual parallel networks. Each parallel network is opened at the bottom of the component that corresponds with the dual component that is at the top of the series network.

Step 2b: Connect the opened parallel connections to the internal nodes of the corresponding series connections.

Step 3: Unroll the network.

### 4.3 Design example

Fig. 5 illustrates both design procedures. The design example consists of a complex differential pull down network used in an OAI22 logic gate (or-and-invert with 2 and 2 inputs). Both design methods produce a network that is fully connected.

In the resulting differential pull down network, both the true and the false of an input signal control a device for every internal node. Consequently independent of the complementary input in the evaluation phase, every internal node is connected to another node. This is either an external node or another internal node, for which both the true and the false of an input signal control a device. As a result independent of the input, every internal node is connected to an external node and the network is fully connected.

## 5 Enhanced fully connected DPDN

The fully connected differential pull down network can be further enhanced to have an evaluation depth that is independent of the discharge events. The evaluation depth is defined as the number of transistors in series between the nodes X or Y to the common ground node Z. A pass-gate is inserted if different discharge paths do not have the same number of transistors. We insert a pass-gate for all the input signals that do not control a transistor in that particular discharge path. A pass-gate, which is built by a parallel combination of two transistors driven by both the signal and its complement, is always open for a complementary input. The enhanced fully connected differential pull down network implementing the AND-NAND function is shown in Fig. 6. The trade-off is an increase in area and total load capacitance.

The introduction of the dummy transistors has two desired effects. First, since the evaluation depth is independent of the discharge event, there is now a constant resistance in the discharge path between outputs X or Y and the common node Z. As a result, each gate has a constant delay as now both the resistance and the capacitance are independent of the inputs. Second, early evaluation has been eliminated. No evaluation will start before all inputs are stable and complementary. As a result, since the logic gate can not produce any anticipated results, the delay of a combination of gates is a constant as each gate evaluates when the input-pair with the longest delay has switched from the 0-0 precharge state to a complementary input.

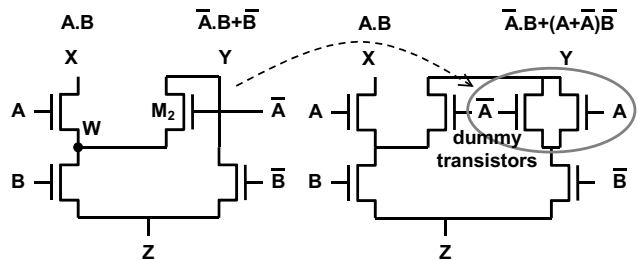

**Fig. 6. Implementation of AND-NAND function with fully connected DPDN (left); and enhanced fully connected DPDN (right).**

## 6 Conclusions

We have illustrated that fully connected differential pull down networks, which are differential transistor networks for which all internal nodes connect to an output node when a complementary input is applied, are necessary to avoid memory effects and to guarantee a constant contribution of the internal nodes into the power consumption of a logic gate. We have presented a design methodology to create fully connected networks. They can be constructed from a given Boolean expression or from a given genuine differential pull down network. Pass gates can be inserted to enhance the fully connected network. The dummy devices ensure a network evaluation depth that is independent of the input signals and eliminate early propagation effects.

## Acknowledgements

The authors would like to acknowledge the support of the National Science Foundation (CCR-0098361).

## 7 References

1. E. Hess, N. Janssen, B. Meyer and T. Schuetze, "Information Leakage Attacks Against Smart Card Implementations of Cryptographic Algorithms and Countermeasures – a Survey," in Proc. of EUROSMART, pp. 55-64, 2000.
2. P. Kocher, J. Jaffe and B. Jun, "Differential Power Analysis," in Proc. of Advances in Cryptology (CRYPTO'99), Lecture Notes in Computer Science, Vol. 1666, pp. 388-397, 1999.
3. M. Renaudin, F. Bouesse, P. Proust, J. Tual, L. Sourgen and F. Germain, "High Security Smart-cards," in Proc. of Design Automation and Test in Europe Conference (DATE 2004), pp. 228-232, 2004.
4. P. Kocher, R. Lee, G. McGraw, A. Raghunathan and S. Ravi, "Security as a New Dimension in Embedded System Design," in Proc. of 41st Design Automation Conference (DAC 2004), pp. 753-760, 2004.
5. S. Chari, C. Jutla, J. Rao and P. Rohatgi, "Towards Sound Approaches to Counteract Power-Analysis Attacks," in Proc. of Advances in Cryptology (CRYPTO'99), Lecture Notes in Computer Science, Vol. 1666, pp. 398-412, 1999.
6. K. Tiri and I. Verbauwhede, "Securing Encryption Algorithms against DPA at the Logic Level: Next Generation Smart Card Technology," in Proc. of Cryptographic Hardware and Embedded Systems (CHES 2003), Lecture Notes in Computer Science, Vol. 2779, pp. 125-136, 2003.
7. J. Fournier, S. Moore, H. Li, R. Mullins and G. Taylor, "Security Evaluation of Asynchronous Circuits," in Proc. of Cryptographic Hardware and Embedded Systems (CHES 2003), Lecture Notes in Computer Science, Vol. 2779, pp. 137-151, 2003.
8. K. Tiri and I. Verbauwhede, "A Logic Level Design Methodology for a Secure DPA Resistant ASIC or FPGA Implementation," in Proc. of Design Automation and Test in Europe Conference (DATE 2004), pp. 246-251, 2004.
9. D. Sokolov, J. Murphy, A. Bystrov and A. Yakovlev, "Improving the Security of Dual-Rail Circuits," in Proc. of Cryptographic Hardware and Embedded Systems (CHES 2004), Lecture Notes in Computer Science, Vol. 3156, pp. 282-297, 2004.
10. K. Tiri, M. Akmal and I. Verbauwhede, "A Dynamic and Differential CMOS Logic with Signal Independent Power Consumption to Withstand Differential Power Analysis on Smart Cards," in Proc. of European Solid-State Circuits Conference (ESSCIRC 2002), pp. 403-406, 2002.
11. F. Mace, F. Standaert, I. Hassoune, J. Legat and J. Quisquater, "A Dynamic Current Mode Logic to Counteract Power Analysis Attacks," in Proc of XIX Conference on Design of Circuits and Integrated Systems (DCIS 2004), 2004.
12. S. Guilley, P. Hoogvorst, Y. Mathieu, R. Pacalet and J. Provost, "CMOS Structures Suitable for Secured Hardware," in Proc. of Design Automation and Test in Europe Conference (DATE 2004), pp. 1414-1415, 2004.
13. M. Allam and M. Elmasry, "Dynamic Current Mode Logic (DyCML): A New Low-Power High-Performance Logic Style," IEEE Journal of Solid-State Circuits, Vol. 36, pp. 550-558, March 2001.
14. L. Heller, W. Griffin, J. Davis and N. Thoma, "Cascode Voltage Switch Logic: A differential CMOS Logic Family," Digest of Technical Papers, IEEE International Solid-State Circuits Conference, pp. 16-17, Feb. 1984.
15. J. Montanaro, et al., "A 160-MHz, 32-b, 0.5-W CMOS RISC Microprocessor," IEEE Journal of Solid-State Circuits, Vol. 31, pp. 1703-1712, Nov. 1996.
16. K. Chu and D. Pulfrey, "Design Procedures for Differential Cascode Voltage Switch Circuits," IEEE Journal of Solid-State Circuits, Vol. 21, pp. 1082-1087, Dec. 1986.
17. J. Rabaey, "Digital Integrated Circuits: A design perspective," Prentice Hall, 1996.